\documentstyle[11pt]{article}
\begin{document}
\begin{titlepage}
\title{\bf Seiberg-Witten Equations on ${\bf R}^{\bf 8}$}
\date{ }
\author{ {\bf Ay\c{s}e H\"{u}meyra Bilge}\\{\small
Department of Mathematics, Institute for Basic Sciences}\\
{\small TUBITAK Marmara Research Center}\\
{\small Gebze-Kocaeli, Turkey}\\
{\footnotesize E.mail : bilge@mam.gov.tr}\\
{\bf Tekin Dereli}\\{\small Department of Physics}\\
{\small Middle East Technical University}\\{\small Ankara, Turkey}
\\{\footnotesize E.mail : tekin@dereli.physics.metu.edu.tr}\\
{\bf \c{S}ahin Ko\c{c}ak}\\{\small Department of Mathematics}\\
{\small Anadolu University}\\{\small Eski\c{s}ehir, Turkey}\\
{\footnotesize E.mail : skocak@vm.baum.anadolu.edu.tr} }
\maketitle
\begin{abstract}
We show that there are no nontrivial solutions of the Seiberg-Witten
equations on $R^8$ with constant standard ${spin}^c$ structure.
\end{abstract} \end{titlepage}

\noindent {\bf 1. Introduction}
\vskip 3mm

The Seiberg-Witten equations are meaningful on any even-dimensional manifold.
To state them, let us recall the general set-up, adopting the terminology of 
the forthcoming book by D.Salamon ([1]).

A $spin^c$-structure on a $2n$-dimensional real inner-product space $V$ is
a pair $(W,\Gamma)$, where $W$ is a $2^n$-dimensional complex Hermitian space
and  $\Gamma:V\rightarrow End(W)$ is a linear map satisfying
$${\Gamma(v)}^*=-\Gamma(v),\qquad {\Gamma(v)}^2=-{\Vert v \Vert}^2$$
for $v\in V$.Globalizing this defines the notion of a $spin^c$-structure
$\Gamma:TX\rightarrow End(W)$ on a $2n$-dimensional (oriented) manifold $X$,
$W$ being a $2^n$-dimensional complex Hermitian vector bundle on $X$. 
Such a structure exists iff $w_2(X)$ has an integral lift. $\Gamma$ extends 
to an isomorphism between the complex Clifford algebra bundle $C^c(TX)$ and 
$End(W)$).
 There is a natural splitting $W=W^+ \oplus W^-$
into the ${\pm}i^n$ eigenspaces of $\Gamma(e_{2n}e_{2n-1}\cdots{e_1})$ 
where $e_1,e_2,\cdots,{e_{2n}}$ is any positively oriented local orthonormal 
frame of $TX$.

The extension of $\Gamma$ to $C_{2}(X)$ gives via the identification of
$\Lambda^{2}(T^{*}X)$ with $C_{2}(X)$ a map
$$\rho:\Lambda^{2}(T^{*}X)\rightarrow End(W)$$
given by
$$\rho(\sum_{i<j}\eta_{ij}e_i^*\wedge e_j^*)=\sum_{i<j}\eta_{ij}\Gamma(e_i)\Gamma(e_j).$$
The bundles $W^\pm$ are  invariant under $\rho(\eta)$ for $\eta\in\Lambda^{2}(T^{*}X)$.
Denote $\rho^\pm (\eta)=\rho(\eta)\vert_{W^\pm}$. The map $\rho$ (and $\rho^\pm$)
extends to
$$\rho:\Lambda^{2}(T^{*}X)\otimes{\bf C}\rightarrow End(W).$$
(If $\eta\in\Lambda^{2}(T^{*}X)\otimes\bf C$ is real-valued then $\rho(\eta)$
is skew-Hermitian and if $\eta$ is imaginary-valued then $\rho(\eta)$ is Hermitian.)

A Hermitian connection $\nabla$ on $W$ is called a $spin^c$ connection 
(compatible with the Levi-Civita connection) if
$$\nabla_v(\Gamma(w)\Phi)=\Gamma(w)\nabla_v\Phi+\Gamma(\nabla_vw)\Phi$$
where $\Phi$ is a spinor (section of $W$),$v$ and $w$ are vector fields on $X$
and $\nabla_vw$ is the Levi-Civita connection on $X$. $\nabla$ preserves
the subbundles $W^\pm$.

There is a principal $Spin^c(2n)=\lbrace e^{i\theta}x\vert \theta\in {\bf R}, x\in Spin(2n) \rbrace
\subset C^c ({\bf R}^{2n})$ bundle  $P$ on $X$ such that $W$ and $TX$ can be
recovered as the associated bundles
$$W=P\times_{Spin^c(2n)}{\bf C}^{2^n}, \qquad TX=P\times_{Ad}{\bf R}^{2n},$$
$Ad$ being the adjoint action of $Spin^c(2n)$ on ${\bf R}^{2n}$.We get then
a complex line bundle $L_{\Gamma}=P\times_{\delta}{\bf C}$ using the map
$\delta:Spin^c(2n)\rightarrow S^1$ given by $\delta ( e^{i\theta}x )
=e^{2i\theta}$.

There is a one-to-one correspondence between $spin^c$ connections on $W$ and
$spin^c(2n)=Lie(Spin^c(2n)=spin(2n)\oplus i{\bf R}$ -valued connection-1-forms $ \hat A\in { \bf A} (P)
\subset \Omega^1 (P,spin^c(2n))$ on $P$.

Now consider the trace-part $A$ of $\hat A$: $A=\frac{1}{2^n}trace(\hat A)$.
This is an imaginary valued 1-form $A\in \Omega^1 (P,i\bf R)$ which is equivariant
and satisfies $$A_p(p\cdot \xi)=\frac{1}{2^n}trace(\xi)$$ for $v\in T_pP,g\in 
Spin^c(2n), \xi \in spin^c(2n)$(where $p\cdot \xi$ is the infinitesimal action).
Denote the set of imaginary valued 1-forms on $P$ satisfying these two 
properties by ${\bf A}(\Gamma)$. There is a one-to-one correspondence between
these 1-forms and $spin^c$ connections on $W$. Denote the connection
corresponding to $A$ by $\nabla_A$. ${\bf A}(\Gamma)$ is an affine
space with parallel vector space $\Omega^1 (X,i\bf R)$. For $A\in {\bf A}(\Gamma)$
the 1-form $2A\in \Omega^1 (P,i\bf R)$ represents a connection on the line
bundle $L_{\Gamma}$. Because of this reason $A$ is called a {\it virtual
connection} on the {\it virtual line bundle} ${L_{\Gamma}^{1/2}}$.
Let $F_A \in\Omega^2 (X,i\bf R)$ denote the curvature of the 1-form $A$.
Finally, let $D_A$ denote the Dirac operator corresponding to $A\in {\bf A}
(\Gamma)$,
$$C^\infty (X,W^+)\rightarrow C^\infty (X,W^-)$$
defined by
$$D_A (\Phi)=\sum_{i=1}^{2n}{\Gamma(e_i){\nabla_{A,e_i}}}(\Phi)$$
where $\Phi \in C^\infty (X,W^+)$ and $e_1,e_2,\cdots,{e_{2n}}$ is any local
orthonormal frame.

The Seiberg-Witten equations can now be expressed as follows. Fix a $spin^c$
structure $\Gamma:TX\rightarrow End(W)$ on $X$ and consider the pairs
$(A,\Phi)\in {{\bf A}(\Gamma) \times C^\infty (X,W^+)}$. The SW-equations
read
$$D_A(\Phi)=0\ \ , \qquad \rho^+(F_A)=(\Phi \Phi^*)_0$$
where $(\Phi \Phi^*)_0 \in C^\infty (X,End(W^+))$
is defined by $(\Phi \Phi^*)(\tau)=<\Phi, \tau> \Phi$ for $\tau \in C^\infty 
(X,W^+)$ and $(\Phi \Phi^*)_0$ is the traceless part of $(\Phi \Phi^*)$.

In dimension $2n=4, \ \rho^+(F_A)= \rho^+(F_A^+) = \rho (F_A^+)$ 
(where $F^+$ is the self-dual part of $F$ and the second equality understood 
in the obvious sense), and therefore self-duality comes intimately into play.
The first problem in dimensions $2n>4$ is that  there is not a generally 
accepted notion of self-duality. Although there are some meaningful definitions 
([2],[3],[4],[5],[6]) (Equivalence of self-duality notions in [2],[3],[5],[6] has 
been shown in [7] , making them more relevant as they separately are), they 
do not assign a well-defined self-dual part to a given 2-form. Even though 
$\rho^+(F_A)$ is still meaningful, it is apparently less important due to the
lack of an intrinsic self-duality of 2-forms in higher dimensions.

The other serious problem in dimensions $2n>4$ is that the SW-equations as 
they are given above are overdetermined. So it is improbable from the outset 
to hope for any solutions. We verify below for $2n=8$ that there aren't 
indeed any solutions. 

In dimension $2n=4$ it is well-known that there are no finite-energy 
solutions ([1]), but otherwise whole classes of solutions are found which 
are related to vortex equations ([8]). In the physically interesting case 
$2n=8$ we will suggest a modified set of equations which is related to
generalized self-duality referred to above. These equations include the 
4-dimensional Seiberg-Witten solutions as special cases.

\vskip 0.5cm
\noindent
{\bf 2. Seiberg-Witten Equations on ${\bf R}^{\bf 8}$}
\vskip .2cm

We fix the constant $spin^c$ structure $\Gamma:{\bf R}^{\bf 8} \longrightarrow 
{\bf C}^{{\bf 16}\times{\bf 16}}$ given by 

$$\Gamma(e_i)=\pmatrix{0&\gamma(e_i)\cr
                        -{\gamma(e_i)}^{*}&0}$$
($e_i,i=1,2,...,8$ being the standard basis for ${\bf R}^{\bf 8}$),where
\vskip .5cm
\noindent
$\gamma(e_1)=\pmatrix{  1&0&0&0&0&0&0&0\cr
                        0&1&0&0&0&0&0&0\cr
                        0&0&1&0&0&0&0&0\cr
                        0&0&0&1&0&0&0&0\cr
                        0&0&0&0&1&0&0&0\cr
                        0&0&0&0&0&1&0&0\cr
                        0&0&0&0&0&0&1&0\cr
                        0&0&0&0&0&0&0&1} $,
$\gamma(e_2)=\pmatrix{  i&0&0&0&0&0&0&0\cr
                        0&-i&0&0&0&0&0&0\cr
                        0&0&-i&0&0&0&0&0\cr
                        0&0&0&-i&0&0&0&0\cr
                        0&0&0&0&i&0&0&0\cr
                        0&0&0&0&0&i&0&0\cr
                        0&0&0&0&0&0&i&0\cr
                        0&0&0&0&0&0&0&-i} $,
\vskip .5cm
$\gamma(e_3)= \pmatrix{ 0&1&0&0&0&0&0&0\cr
                        -1&0&0&0&0&0&0&0\cr
                         0&0&0&0&-1&0&0&0\cr
                         0&0&0&0&0&-1&0&0\cr
                         0&0&1&0&0&0&0&0\cr
                         0&0&0&1&0&0&0&0\cr
                         0&0&0&0&0&0&0&1\cr
                         0&0&0&0&0&0&-1&0} $,

\vskip .5cm

$\gamma(e_4)= \pmatrix{ 0&i&0&0&0&0&0&0\cr
                         i&0&0&0&0&0&0&0\cr
                         0&0&0&0&-i&0&0&0\cr
                         0&0&0&0&0&-i&0&0\cr
                         0&0&-i&0&0&0&0&0\cr
                         0&0&0&-i&0&0&0&0\cr
                         0&0&0&0&0&0&0&i\cr
                         0&0&0&0&0&0&i&0} $,
\vskip .5cm

$\gamma(e_5)= \pmatrix{ 0&0&1&0&0&0&0&0\cr
                        0&0&0&0&1&0&0&0\cr
                        -1&0&0&0&0&0&0&0\cr
                         0&0&0&0&0&0&-1&0\cr
                         0&-1&0&0&0&0&0&0\cr
                         0&0&0&0&0&0&0&-1\cr
                         0&0&0&1&0&0&0&0\cr
                         0&0&0&0&0&1&0&0} $,
\vskip .5cm
$\gamma(e_6)= \pmatrix{ 0&0&i&0&0&0&0&0\cr
                        0&0&0&0&i&0&0&0\cr
                         i&0&0&0&0&0&0&0\cr
                         0&0&0&0&0&0&-i&0\cr
                         0&i&0&0&0&0&0&0\cr
                         0&0&0&0&0&0&0&-i\cr
                         0&0&0&-i&0&0&0&0\cr
                         0&0&0&0&0&-i&0&0} $,

$\gamma(e_7)= \pmatrix{ 0&0&0&1&0&0&0&0\cr
                        0&0&0&0&0&1&0&0\cr
                         0&0&0&0&0&0&1&0\cr
                         -1&0&0&0&0&0&0&0\cr
                         0&0&0&0&0&0&0&1\cr
                         0&-1&0&0&0&0&0&0\cr
                         0&0&-1&0&0&0&0&0\cr
                         0&0&0&0&-1&0&0&0} $,
\vskip .5cm
$\gamma(e_8)= \pmatrix{ 0&0&0&i&0&0&0&0\cr
                        0&0&0&0&0&i&0&0\cr
                         0&0&0&0&0&0&i&0\cr
                         i&0&0&0&0&0&0&0\cr
                         0&0&0&0&0&0&0&i\cr
                         0&i&0&0&0&0&0&0\cr
                         0&0&i&0&0&0&0&0\cr
                         0&0&0&0&i&0&0&0} $.

\vskip .5cm
(We obtain this $spin^c$ structure from the well-known isomorphism of the
complex Clifford algebra $C^c({\bf R}^{2n})$ with $End(\Lambda ^*{{\bf C}^
n})$.)

In our case $X={\bf R}^8, W={\bf R}^8 \times {\bf C}^{16}, 
W^\pm={\bf R}^8 \times {\bf C}^8$ and 
$L_\Gamma = {L_\Gamma}^{1/2}={\bf R}^8 \times
 {\bf C}$.
Consider the connection 1-form 
$$A=\sum_{i=1}^8 A_i dx_i \in {\Omega}^1 ({\bf R}^8,i{\bf R})$$
on the line bundle ${\bf R}^8 \times {\bf C}$. Its curvature is given by
$$F_A=\sum_{i<j}F_{ij} dx_i \wedge dx_j \in {\Omega}^2 ({\bf R}^8,i{\bf R})$$
where 
$F_{ij}=\frac {\partial A_j}{\partial x_i} - \frac {\partial A_i}{\partial 
x_j}$.
The $spin^c$ connection $\nabla= \nabla_A$ on $W^+$ is given by
$$\nabla_i{\Phi}= \frac {\partial \Phi}{\partial x_i}+ A_i \Phi$$
($i=1,...,8$) where $\Phi : {\bf R}^8 \rightarrow {\bf C}^8$.

$$\rho^+:\Lambda^{2}(T^{*}X)\otimes{\bf C}\rightarrow End(W^+)$$
is given by

\noindent
$\rho^+(F_A)= \pmatrix{ G_{11}&G_{12}&G_{13}&G_{14}&G_{15}&G_{16}&G_{17}&0\cr
         \bar G_{12}&G_{22}&G_{23}&G_{24}&G_{25}&G_{26}&0&G_{17}\cr
   \bar G_{13}&\bar G_{23}&G_{33}&G_{34}&G_{35}&0&G_{26}&-G_{16}\cr
  \bar G_{14}&\bar G_{24}&\bar G_{34}&G_{44}&0&G_{35}&-G_{25}&G_{15}\cr
  \bar G_{15}&\bar G_{25}&\bar G_{35}&0&-G_{44}&G_{34}&-G_{24}&G_{14}\cr
  \bar G_{16}&\bar G_{26}&0&\bar G_{35}&\bar G_{34}&-G_{33}&G_{23}&-G_{13}\cr
  \bar G_{17}&0&\bar G_{26}&-\bar G_{25}&-\bar G_{24}&\bar G_{23}&-G_{22}&G_{12}\cr
  0&\bar G_{17}&-\bar G_{16}&\bar G_{15}&\bar G_{14}&-\bar G_{13}&\bar G_{12}&-G_{11}},$

where

$$\begin{array}{ll}
          G_{11}={iF_{12}+iF_{34}+iF_{56}+iF_{78}},&
          G_{12}={F_{13}+iF_{14}+iF_{23}-F_{24}},\cr
          G_{13}={F_{15}+iF_{16}+iF_{25}-F_{26}},&
          G_{14}={F_{17}+iF_{18}+iF_{27}-F_{28}},\cr
          G_{15}={F_{35}+iF_{36}+iF_{45}-F_{46}},&
          G_{16}={F_{37}+iF_{38}+iF_{47}-F_{48}},\cr
          G_{17}={F_{57}+iF_{58}+iF_{67}-F_{68}},&
          G_{22}={-iF_{12}-iF_{34}+iF_{56}+iF_{78}},\cr
          G_{23}={-F_{35}-iF_{36}+iF_{45}-F_{46}},&
          G_{24}={-F_{37}-iF_{38}+iF_{47}-F_{48}},\cr
          G_{25}={F_{15}+iF_{16}-iF_{25}+F_{26}},& 
          G_{26}={F_{17}+iF_{18}-iF_{27}+F_{28}},\cr
          G_{33}={-iF_{12}+iF_{34}-iF_{56}+iF_{78}},&
          G_{34}={-F_{57}-iF_{58}+iF_{67}-F_{68}},\cr
          G_{35}={-F_{13}-iF_{14}+iF_{23}-F_{24}},&
          G_{44}={-iF_{12}+iF_{34}+iF_{56}-iF_{78}.}
\end{array}
$$
For $\Phi=(\phi_1,\phi_2,...,\phi_8) \in C^\infty(X,W^+)=C^\infty ({\bf R}^8,
{\bf R}^8\times {\bf C}^8)$,
$$(\Phi {\Phi}^*)_0=
\pmatrix {\phi_1 \bar{\phi_1}-{1/8}\sum{\phi_i \bar \phi_i}&\phi_1 \bar{\phi_2}&.&.&.&\phi_1 \bar{\phi_8}\cr
                           \phi_2 \bar{\phi_1}&\phi_2 \bar{\phi_2}-{1/8}\sum{\phi_i \bar \phi_i}&.&.&.&\phi_2 \bar{\phi_8}\cr
                          .&.&.&.&.&.\cr .&.&.&.&.&.\cr .&.&.&.&.&.\cr
                          \phi_8 \bar{\phi_1}&\phi_8 \bar{\phi_2}&.&.&.&\phi_8 \bar{\phi_8}-{1/8}\sum{\phi_i \bar \phi_i}}.
$$

It was remarked by Salamon([1],p.187) that $\rho^+(F_A)=0$ implies $F_A=0$.
(i.e.reducible solutions of 8-dim. SW-equations are flat.)\\
It can be explicitly verified that all solutions are reducible and flat:\\
{\bf Proposition:} There are no nontrivial solutions of the Seiberg-Witten
equations on ${\bf R}^8$ with constant standard ${spin}^c$ structure,i.e.
\\
$\rho^+(F_A)=(\Phi \Phi^*)_0$ (alone) implies $F_A=0$ and $\Phi=0$.
\\
Proof: Trivial but tedious manipulation with the linear system.\\

\vskip 4mm
\noindent {\bf Acknowledgement}
\vskip 3mm

The above  work is based on a talk given at the 
{\bf 5th G\"{o}kova Geometry-Topology Conference} held at Akyaka-Mu\u{g}la, 
Turkey during May, 1996.

\newpage
\noindent {\bf References}
\vskip 5mm
\begin{description}
\item{[1]} D.Salamon, {\bf Spin Geometry and Seiberg-Witten Invariants}
(April 1996 version)(to appear).
\item {[2]} A.Trautman, Int.J.Theo.Phys.16(1977)561.
\item {[3]} D.H.Tchrakian, J.Math.Phys.21(1980)166.
\item {[4]} E.Corrigan, C.Devchand, D.B.Fairlie, J.Nuyts, Nucl.Phys.B 214(1983)
452.
\item {[5]} B.Grossman, T.W.Kephart, J.D.Stasheff, Commun.Math.Phys.96(1984)431
Erratum:ibid,100(1985)311.
\item {[6]} A.H.Bilge, T.Dereli, \c{S}.Ko\c{c}ak, Lett.Math.Phys.36(1996)301.
\item {[7]} A.H.Bilge, Self-duality in dimensions $2n>4$, dg-ga/9604002.
\item {[8]} C.Taubes, SW $\rightarrow$ Gr, J. of the A.M.S.9,3(1996).
\end{description}

\end{document}